\documentclass[11pt]{article}

\usepackage{amsfonts}
\usepackage{amsmath}
\usepackage{amssymb}
\usepackage{latexsym}
\usepackage{times}
\usepackage{epsfig}
\usepackage{multirow}
\usepackage{hyperref}
\usepackage{url}
\newcommand{\ket}[1]{|\, #1\rangle}
\newcommand{\bra}[1]{\langle #1\,|}

\newcommand\beq{\begin{equation}}
\newcommand\eeq{\end{equation}}
\newcommand\bea{\begin{eqnarray}}
\newcommand\eea{\end{eqnarray}}

 \def\squarebox#1{\hbox
to #1{\hfill\vbox to #1{\vfill}}}
\def\qed{\hspace*{\fill}\vbox{\hrule\hbox{\vrule\squarebox{.667em}\vrule}\hrule}
} 
\newenvironment{proof}{\begin{trivlist}\item[]{\bf Proof:}}{\qed
\end{trivlist}}

\newcommand{\ba}{\begin{array}}
\newcommand{\ea}{\end{array}}

\newtheorem{theo}{Theorem}

\newtheorem{defi}{Definition}

\newtheorem{lem}{Lemma}

\newtheorem{propo}{Proposition}

\begin{document}

\author{Barbara M. Terhal\footnote{ITFA, Universiteit van Amsterdam, Valckenierstraat 65, 1018 XE, Amsterdam, The Netherlands.}'
\footnote{IBM Watson Research Center, P.O. Box 218, Yorktown Heights, NY 10598, USA.}
\;and Guido Burkard\footnotemark[2]}


\title{Fault-Tolerant Quantum Computation For Local Non-Markovian Noise}

\date{}


\maketitle

\begin{abstract}
We derive a threshold result for fault-tolerant quantum computation for local non-Markovian noise models.
The role of error amplitude in our analysis is played by the product of the elementary
gate time $t_0$ and the spectral width of the interaction Hamiltonian between system and bath. We discuss
extensions of our model and the applicability of our analysis.
\end{abstract}

\section{Introduction}

Whether or not quantum computing will become reality will at some
point depend on whether we can implement quantum computation
fault-tolerantly. This would imply that even though the quantum
circuitry and storage are faulty, it is possible by error-correction
to perform errorfree quantum computation for an
unlimited amount of time while incurring an overhead that is
polylogarithmic in time and space, see \cite{shor-faulttol},
\cite{AB-faulttol}, \cite{AB-ftsiam}, \cite{klz-faulttol},
\cite{gottesman-faulttol}, \cite{preskill-faulttol} and
\cite{KLZ-res}. For this `software' solution that uses
concatenated coding techniques, an error probability threshold of
the order of $10^{-4}-10^{-6}$ per qubit per clock-cycle has been
given for the simplest error models, meaning that for an error
probability below this threshold fault-tolerant quantum
computation is possible. These estimates heavily depend on error
modelling, the efficiency of the error-correcting circuits, and
the codes that are used.  Different and potentially better
estimates are possible, see for example \cite{steane-overhead},
\cite{DB-pureft} and \cite{knill-newft}. Another solution to the
fault-tolerance problem proposed by Kitaev is to make the hardware
intrinsically fault-tolerant by using topological degrees of
freedom such as anyonic excitations as qubits \cite{kitaev-top}.

In Refs.~\cite{klz-faulttol} and \cite{AB-ftsiam} the threshold result for fault-tolerance is derived
for various error models, including ones with exponentially decaying correlations.
However, this general model of exponentially decaying correlations does not make direct contact
with a detailed physical model of decoherence. Such a physical
model of decoherence starts from a Hamiltonian description involving the environmental
degrees of freedom and the computer `system' degrees of freedom.

Starting from such a Hamiltonian picture it was argued in a paper by Alicki {\em et al.}
\cite{alicki+-faulttol} that fault-tolerant quantum computation may not be
possible when the environment of the quantum computer has a long-time
memory.

In this paper we carry out a detailed threshold analysis for some non-Markovian error models. Our findings are not
in agreement with the views put forward in the paper by Alicki {\em et al.}, that is, we can derive a threshold result
in the non-Markovian regime if we make certain reasonable assumptions about the spatial structure and interaction amongst the
environments of the qubits. The results of our paper and the previous results in the literature are summarized
in Section \ref{soa} of this paper. In section \ref{decohmodel} we introduce our notation and our assumptions
on the decoherence model. In section \ref{just} we introduce our measure of error or decoherence strength which
we motivate with a small example. Then in section \ref{emt} we prove some simple lemmas that will be used in the
fault-tolerance analysis and in section \ref{overall} we discuss the overall picture of a fault-tolerance derivation,
in particular the parts of this derivation that do not depend on the decoherence model.
Then in Section \ref{threshsec} we fill in the technical details to obtain the threshold result expressed in Theorem 1.
In Section \ref{bathext} we generalize our decoherence model to incorporate more relaxed conditions on the
spatial structure of the bath and we discuss further possible extensions.
In Section \ref{soa} we give an overview of all known fault-tolerance results including ours and in the
last section \ref{strength} we discuss several physical systems in which our analysis may be applicable.

\subsection{Notation and Explanation of the Decoherence Model}
\label{decohmodel}

We use the following operator norm: $||A||=\max_{||\psi||=1} ||A
\ket{\psi}||$ where $||\,\ket{\psi}|| \equiv
||\psi||=\sqrt{\bra{\psi}\psi\rangle}$. The following
properties will be used:
 $||A+B|| \leq ||A||+||B||$, $||U||=1$ if $U$ is unitary, and $||A B|| \leq ||A|| \,||B||$.
An operator $H$ that acts on system qubit $i$ or qubits $i$ and $j$ (and
potentially another quantum system) is denoted as
$H[\texttt{q}_i]$ or $H[\texttt{q}_i,\texttt{q}_j]$. A unitary
evolution for the time-interval $t$ to $t+t_0$ is denoted as
$U(t+t_0,t)$. $t_0$ is the time it takes to do an elementary (one or two qubit) gate. The identity operator
is denoted as ${\bf I}$ and ${\bf e}$ denotes the base of the natural logarithm. We will
also use the trace-norm denoted by $||A||_1={\rm Tr}
\sqrt{A^{\dagger}A}$ and the classical variation distance between
probability distributions $\mathbb{P}$ and $\mathbb{Q}$: $||\mathbb{P}-\mathbb{Q}||_1=\sum_i
|\mathbb{P}(i)-\mathbb{Q}(i)|$.

\begin{figure}[htb]
\begin{center}
\epsfxsize=15cm \epsffile{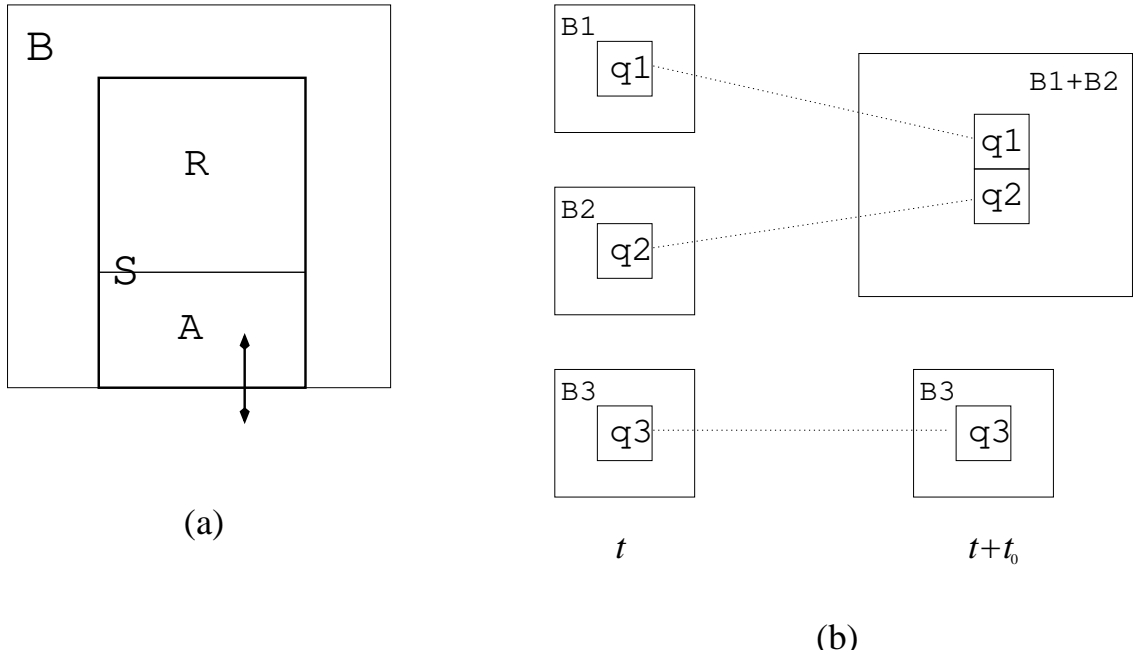} \caption{Schematic
representation of the model.  (a) The system \texttt{S} consists
of a register \texttt{R} of qubits plus ancillas \texttt{A} that
can be reset during the computation.  The system \texttt{S} is
coupled to the environment, or bath,  \texttt{B}. (b) The
decoherence model.  Each qubit $\texttt{q}_i$ is coupled to an
individual bath $\texttt{B}_i$.  When two qubits interact, they
may interact with one common bath.}
\label{figschem}
\end{center}
\end{figure}

The following assumptions have been shown to be necessary 
for fault-tolerance and thus
we keep these assumptions in our analysis:
\begin{itemize}
\item It is possible to operate gates on different qubits in parallel.
\item We have fresh ancilla qubits to our disposal. These ancilla qubits are prepared off-line
in the exact computational state $\ket{00 \ldots 0}$ and they can be used in the circuit  
when necessary. They function as a
heat-sink which removes entropy from the computation.
\end{itemize}.


In Figure \ref{figschem}(a) three types of
quantum systems are sketched that differ in function and in the amount of control that we can exert over them.
First, there is \texttt{R}, for quantum Registers, that we can control and use for our computation.
Secondly, there is \texttt{A}, for Ancillas, which are used for error-correction and fault-tolerant
gate construction 
during the computation. The systems \texttt{R} and \texttt{A} taken together are denoted as \texttt{S} for
System of which single qubits are denoted by the letter $\texttt{q}$.
Clean ancilla registers set to $\ket{00 \ldots 0}$ are added during the computation and can be removed
after having interacted with (1) other parts of the system $\texttt{S}$ by
error-correcting procedures and (2) the bath \texttt{B} according
to some fixed interaction Hamiltonian.  

We will assume that the third system, the bath \texttt{B}, which interacts with system and  
ancillas has a local structure, illustrated in Fig.~\ref{figschem}(b). 
We will generalize this model in Section \ref{bathext}. Every
qubit ($\texttt{q}_1,\texttt{q}_2,\texttt{q}_3$...) of the system
has its own bath $(\texttt{B}_1,\texttt{B}_2,\texttt{B}_3..)$.
Only during the time when two qubits interact their baths   
($\texttt{B}_1$ and $\texttt{B}_2$ in the figure) can interact. The
idea behind this modelling is that the bath is localized in space,
i.e. is associated with the place where the qubit is stored. But when qubits
interact, they need to be brought together and so they may share a
common bath. In the picture $\texttt{B}_1+ \texttt{B}_2$ at time
$t+t_0$ are suggested to be the same baths that qubits $\texttt{q}_1$ and
$\texttt{q}_2$ interacted with at time $t$, but in general they
may also be different baths. For example,  when qubits  
$\texttt{q}_1$ and $\texttt{q}_2$ have to be moved in order to
interact, they may see a partially new environment at time $t+t_0$.
This distinction will not be important in our analysis.

Most importantly, in this model, each bath can have an arbitrarily 
long memory; at no point in our derivation will we make a Markovian assumption. This implies that, for example,
 the bath $\texttt{B}_1$ may contain information about qubit $\texttt{q}_1$ at time $t$, then interact with bath
 $\texttt{B}_2$ at
 time $t+t_0$ and pass this information on to bath $\texttt{B}_2$ etc.
The interaction Hamiltonian of a single qubit $\texttt{q}_i$ of the system (\texttt{R} or \texttt{A}) with
the bath is given by
\beq
H_{\texttt{SB}}[\texttt{q}_i]=\sum_k \sigma_k[\texttt{q}_i]\otimes A_k.
\eeq
with the Pauli-matrices $\sigma_k$ acting on qubit $\texttt{q}_i$ and $A_k$ is some Hermitian operator
on the bath of the qubit $\texttt{q}_i$ which is not equal to the identity ${\bf I}$. During a
two qubit-gate both qubits may interact with both baths. For simplicity (see footnote \cite{fn1})
we assume that the interaction is of the form
\beq
H_{\texttt{SB}}[\texttt{q}_i,\texttt{q}_j]=H_{\texttt{SB}}[\texttt{q}_i]+H_{\texttt{SB}}[\texttt{q}_j],
\eeq
where the bath part of each $H_{\texttt{SB}}[\texttt{q}_i]$ is an operator on the joint bath of qubits $\texttt{q}_i$
and $\texttt{q}_j$. We do not care about the time-evolution of the baths except that it has to obey the ``local bath assumption'', i.e.
noninteracting qubits have noninteracting baths. The system (register and ancilla)  
evolution $H_{\texttt{RA}}(t)$ is time-dependent
and represents the fault-tolerant quantum circuit that we want to implement.
This evolution is built from a sequence
of one and two qubit gates and, as was said before, 
$t_0$ is the time it takes to perform any such gate.

\subsection{Measure of Decoherence Strength}
\label{just}

Our results will depend on the strength of the coupling
Hamiltonian $H_{\texttt{SB}}[\texttt{q}
_i]$. There is an
additional freedom in determining $H_{\texttt{SB}}[\texttt{q}_i]$,
namely we can always add a term $ \alpha {\bf
I}_{\texttt{S}}[\texttt{q}_i] \otimes {\bf I}_{\texttt{B}}$ where
$\alpha$ is an arbitrary real constant and ${\bf I}$ is the
identity operator. This is possible since it merely shifts the
spectrum (see footnote \cite{fn2}). Let
$\mu_i$ be the eigenvalues of $H_{\texttt{SB}}$. With this freedom we see that \beq
\min_{\alpha}||H_{\texttt{SB}}[\texttt{q}_i]+\alpha {\bf
I}_{\texttt{S}}[\texttt{q}_i] \otimes {\bf
I}_{\texttt{B}}||=(\mu_{{\rm max}}-\mu_{{\rm min}})/2
\equiv \Delta_{\texttt{SB}}[\texttt{q}_i], \label{minnorm} \eeq
the spectral width of the interaction Hamiltonian (divided by 2). Our analysis
will apply to physical systems where one can bound \beq
\forall\;\texttt{q}_i \in \texttt{S},\;
\Delta_{\texttt{SB}}[\texttt{q}_i] \leq \lambda_0. \eeq where
$\lambda_0$ is a small constant which will enter the threshold
result, Theorem \ref{thresh1}, together with $t_0$, the
fundamental gate time. In what follows we will denote $\Delta_{\texttt{SB}}[\texttt{q}_i]$ as
$\Delta_{\texttt{SB}}$ or $\Delta$ assuming that the spectral width is the same
for each qubit in $\texttt{S}$.

We justify the use of this norm in the following way. Consider a
single qubit coupled to a bath such that both bath and system
Hamiltonians are zero but there exists nonzero coupling. To what
extent will an arbitrary initial state of qubit and bath change
under this interaction? We can consider the minimum
fidelity of an initial state $\psi_{\texttt{SB}}(0)$ with the evolved state at time $t$: \beq
F_{min}(t)=\min_{\psi(0)}|\bra{\psi(t)}\psi(0)\rangle|. \eeq  For small
times $t$ such that $\Delta_{\texttt{SB}} t \leq
\pi/2$ the minimum fidelity can be achieved by taking
$\ket{\psi(0)}=\frac{1}{\sqrt{2}}( \ket{\psi_{{\rm
max}}}+\ket{\psi_{{\rm min}}})$ where $\ket{\psi_{{\rm max/min}}}$
are the eigenvectors of $H_{\texttt{SB}}$ with largest and
smallest eigenvalues. Then we have \beq F_{min}(t)=\cos(\Delta t)
\approx 1-\Delta^2 t^2/2+O((\Delta t)^4). \eeq
Note that this fidelity decay {\em includes} the effects on the bath. For this reason this fidelity decay
overestimates the effects of decoherence, in other words $F(\rho_{\texttt{S}}(t),\rho_{\texttt{S}}(0)) \geq F_{min}$.

One may compare this fidelity decay with that of other decoherence processes,
for example the depolarizing channel ${\cal E}$ with depolarizing probability $p$.
For such a channel we have
$F(\ket{\psi}_{\texttt{S}},{\cal E}(\ket{\psi}\bra{\psi})_{\texttt{S}})=\sqrt{1-\frac{p}{2}}$
\cite{book-nielsen&chuang}. Thus, loosely speaking, $\Delta t$ could be interpreted as an error amplitude
whose square is an error probability.

Thus, this brief analysis shows that for some initial states $\psi_{\texttt{SB}}(0)$ the
norm of the interaction Hamiltonian measures exactly how the state changes
due to the interaction. Since our environment is non-Markovian we cannot
exclude such bad initial states, in other words we cannot assume that the
decoherence is just due to the interactive evolution of an initially
unentangled bath and system.


\subsection{Error Modelling Tools}
\label{emt}
The following simple lemma will be used repeatedly in this paper:

\begin{lem} Let a unitary transformation ${\bf U}=U_n \ldots U_1$ where $U_i=G_i+B_i$ and the operator $G_i$ and $B_i$ are not necessarily unitary. Let
${\bf U}={\bf B}+{\bf G}$ where we define ${\bf B}$ to be the sum
of terms which contains at least $k$ factors $B_i$. Let  $||B_i|| \leq \epsilon$ and thus $||G_i|| \leq 1+\epsilon$.
We have
\beq
||{\bf B}|| \leq {n \choose k} \epsilon^k (1+\epsilon)^{n-k}.
\eeq
If $G_i$ is unitary, we have   
\beq
||{\bf B}|| \leq {n \choose k} \epsilon^k.
\eeq
\label{blemma1}
\end{lem}

\begin{proof} We can think about ${\bf U}$ as a binary tree of depth $n$ such that the children of each node are
labelled with $G_i$ or $B_i$ at depth $i$. We prune the tree in the
following way; when a branch has $k$ factors $B_i$ in its path, we terminate this whole branch with the remaining $U_n \ldots U_m$. The sum of these terminated
branches is ${\bf B}$. ${\bf B}$ can be bounded by observing that there are ${n \choose k}$ terminated branches each of which have norm at most
$||B_i||^k ||G_i||^{n-k}$ (since each branch is a sequence of $G_i$ transformations interspersed with $k$ $B_i$ transformations followed by unitary transformations).
\end{proof}

It is easy to prove the following (see also Ref. \cite{KLV-noise})

\begin{lem}
Consider a time-interval $[t,t+t_0]$ and a single qubit $\texttt{q} \in \texttt{S}$
which does not interact with any other qubit in \texttt{S} at that time. The time-evolution for this
qubit is given by some unitary evolution $U[\texttt{q}]$
involving its bath $\texttt{B}$. Let $U_0[\texttt{q}]=U_{\texttt{S}}[\texttt{q}]\otimes U_{\texttt{B}}$ be
the free uncoupled evolution for this qubit. We can write
\beq
U[\texttt{q}]=U_0[\texttt{q}]+E[\texttt{q}],
\eeq
where $E[\texttt{q}]$ is a fault-operator with norm
\beq
||E[\texttt{q}]|| \leq t_0 ||H_{\texttt{SB}}[\texttt{q}]||= t_0 \Delta_{\texttt{SB}}[\texttt{q}] \leq  t_0 \lambda_0.
\eeq
\label{ebound}
\end{lem}

\begin{proof}
We drop writing the dependence on qubit $\texttt{q}$ for the proof. For the qubit evolution in the interval, using the Trotter expansion we can write
\beq
U=\lim_{n \rightarrow \infty}\Pi_{m=1}^n (U_{\texttt{S}}^{t_m} U_{\texttt{SB}}^{t_m} U_{\texttt{B}}^{t_m}).
\eeq
where $U_{\texttt{K}}^{t_m}$ is the time-evolution for $\texttt{K}=\texttt{S}, \texttt{B}$ or coupling $\texttt{SB}$
during the time-interval $t_m$ of length $t_0/n$. Now in this expansion we may write
$U_{\texttt{SB}}^{t_m}={\bf I}-i H_{\texttt{SB}}t_0/n+O(\frac{t_0^2}{n^2})$ and omit these higher order terms.
Let us call $G_m=U_{\texttt{S}}^{t_m}U_{\texttt{B}}^{t_m}$ and $B_m=-i\frac{t_0}{n} U_{\texttt{S}}^{t_m} H_{\texttt{SB}} U_{\texttt{B}}^{t_m}$ as in
Lemma \ref{blemma1}. We thus have $||B_m|| \leq \frac{t_0}{n} ||H_{\texttt{SB}}||$.
Note that $G_m$ is unitary and we have a binary tree
of depth $n \rightarrow \infty$ and can use Lemma \ref{blemma1} with $k=1$. This
gives
\beq
||E||=||{\bf B}|| \leq t_0 ||H_{\texttt{SB}}||.
\eeq
\end{proof}

A similar statement holds when we consider the evolution of two interacting qubits. We have that
\beq
U_{SB}[\texttt{q}_i,\texttt{q}_j]=U_0[\texttt{q}_i,\texttt{q}_j]+E[\texttt{q}_i,\texttt{q}_j],
\eeq
where $||E[\texttt{q}_i,\texttt{q}_j]|| \leq 2 t_0 \Delta_{SB}[\texttt{q}] \leq 2 t_0 \lambda_0$.

\subsection{Overall Perspective: good and bad fault-paths}
\label{overall}

Since the bath may retain information about the time-evolution and error processes for arbitrary long times we cannot describe the
decoherence process by sequences of superoperators on the system qubits. Instead, there is a single superoperator
for the entire computation that is obtained by tracing over the bath at the end of the computation.
Thus in our analysis we will consider the entire {\em unitary} evolution of system, bath and ancillas. At time $t=0$ bath and ancilla
and system are uncoupled and we may always purify the bath, i.e., find a pure state in a larger
bath Hilbert space which, when the extra Hilbert space is traced out, yields the
desired mixed state.  
We can then assume a pure initial product state for the combined system and bath, \texttt{SB}.
The unitary evolution of the computation consists of a sequence and/or parallel application of
the unitary gates $U[\texttt{q}_i,\texttt{q}_j](t+t_0,t)$ and $U[\texttt{q}_i](t+t_0,t)$. Each such gate, say
for two qubits, can be written as a sum of a error-free evolution $U_0[\texttt{q}_i,\texttt{q}_j](t+t_0,t)$ and
a fault term $E[\texttt{q}_i,\texttt{q}_j]$. Therefore the entire computation can be written as a sum over
{\em fault-paths}, that it, a sum of sequences of unitary error-free operators interspersed with fault operators.
This is very similar as in the fault-tolerance analysis for Markovian error models, where the superoperator
during each gate-time $t_0$ can be expanded in a error-free evolution and an erroneous evolution so that
the entire superoperator for the circuit is a sum over fault-paths.

The main idea behind the threshold result for fault-tolerance is then as follows, see \cite{AB-ftsiam}.
There are \emph{good} fault-paths with so called \emph{sparse} numbers of faults which keep being corrected during the computation and
which lead to (approximately) correct answers of the computation. And there are \emph{bad} fault-paths which
contain too many faults to be corrected and imply a {\em crash} of the quantum computer.

Now the goal of our fault-tolerance derivation which is completely analogous in structure as the one in
\cite{AB-ftsiam} is to show the following:
\begin{enumerate}
\item[{\sf A}] Sparse fault-paths lead to sparse errors in the
computation. This fact uses the formal distinction between faults
that occur during the computation and the effects of these faults,
the errors, that arise due the subsequent evolution which can
spread the faults. The fact that sparse fault-paths give rise to
sparse errors is due to fundamental properties of fault-tolerant
error-correcting circuitry, namely that there exists
error-correcting codes and procedures that do not spread faults
too much. It is independent of the choice of decoherence model,
and can be applied to any model where one can make an expansion
into fault-paths. See Lemma \ref{codelem}.
\item[{\sf B}] Sparse
errors give good final answers. This is a technical result whose
derivation may differ slightly in one or the other decoherence
model, but which is intuitively sound for all possible decoherence
models. See Lemma \ref{dev}. \item[{\sf C}] The norm of the
operator corresponding to all bad non-sparse fault-paths is
``small''. This result depends crucially on the decoherence model
that is chosen, in particular the spatial or temporal correlations
that are allowed. Secondly, it depends on the strength of the
errors, that is, only for small enough strength below some
threshold value will the norm of the bad fault-path operator get
small. See Section \ref{secC}. \item[{\sf A,B,C}$\Rightarrow$]
When the \emph{bad} operator norm is small, the answer of the
computation is close to what the good fault-path operator yields
which is the correct answer according to item {\sf B}. See Lemma
\ref{dev} and Theorem \ref{thresh1}.
\end{enumerate}

Another small comment about our model is the following:
In the usual model for error-correction (see Ref. \cite{preskill-faulttol} in \cite{book-LPS}),
measurements are performed to determine the error-syndrome
or the correct preparation of the ancilla states. Since
we prefer to view the entire computation as a unitary process, we may replace
these measurements by coherent quantum operations. In the error-correction
with measurement procedures it is assumed that faults can occur in the
measurement itself or in the quantum gate that is performed that depends on this
measurement record, but the measurement record by itself is stable since it is classical.
If we replace measuring by coherent action for technical reasons in this
 derivation, it is then fair to assume that the qubit that carries
 the measurement record is errorfree, in other words does no longer interact with a bath.
 This modelling basically allows the standard fault-tolerance results in item {\sf A} expressed in
  Lemma \ref{codelem}, to carry over in the simplest way to our model.

\section{Threshold result}
\label{threshsec}

\subsection{Nomenclature}
\label{nomen}

\begin{figure}[htb]
\begin{center}
\epsfxsize=10cm \epsffile{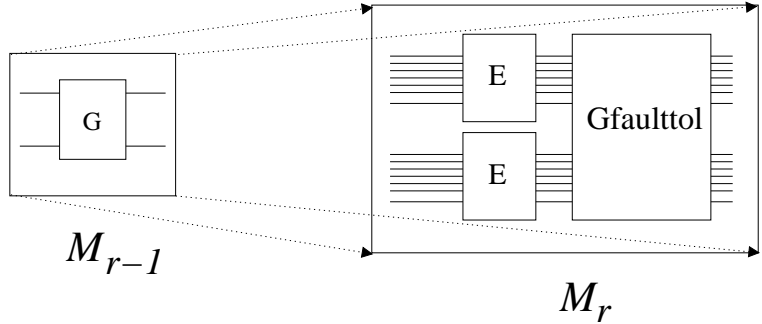} \caption{Every single or
two-qubit gate G in the circuit $M_{r-1}$ gets replaced by an
error-correcting procedure E followed by a fault-tolerant
implementation of G, Gfaulttol (possibly involving ancillas).}
\label{concat}
\end{center}
\end{figure}

Let the basic errorfree quantum circuit denoted by $M_0$ consist of $N$ {\em locations} \cite{AB-ftsiam}.
Each location is given by a triple $(\{\texttt{q}\},{\rm G},t)$ where $\{\texttt{q}\}$ denotes the qubits
(one or two at most) involved in some gate G (G could be ${\bf I}$) at time $t$ in the quantum circuit. In the following, $E[i]$ or
$U[i]$ will denote operators that involve location $i$, i.e. if $\texttt{q}_1$ and $\texttt{q}_2$ interact at
location $i$ we will write $U_0[\texttt{q}_1,\texttt{q}_2]=U_0[i]$ instead of enumerating the qubits.
For fault-tolerance one constructs a
family of circuits $M_r$ by concatenation. That is, we fix a computation code $C$ (see definition 15 in
Ref. \cite{AB-ftsiam}), for example a CSS code,  encoding one qubit into (say) $m$ qubits \cite{fn3}.
We obtain the circuit $M_r$ by replacing each location in the circuit $M_0$
by a block of encoded qubits to which we apply an error-correcting procedure followed by a fault-tolerant implementation of G, see Fig. \ref{concat}.
Repeated substitution will gives us a circuit $M_r$ at concatenation level $r$.

Essential are the following definitions and a lemma taken from Ref.
\cite{AB-ftsiam} which define sparseness of a set of locations
and error-spread of a code:\\

{\bf Definitions from Ref. \cite{AB-ftsiam}:}
\begin{itemize}
{\it \item A set of qubits in $M_r$ is called an $s$-block if they originate from 1 qubit in $M_{r-s}$.
A $s$-working period in $M_r$ is a time-interval which originates from
one time-step in $M_{r-s}$. 
A $s$-rectangle in $M_r$ is a set of locations that originate from one location in $M_{r-s}$.
\item
Let $B$ be a set of $r$-blocks in the circuit $M_r$. An $(r,1)$-sparse set of qubits
$A$ in $B$ is a set of qubits in which for every $r$-block in $B$, there is
at most one $(r-1)$-block such that the set $A$ in this block is not $(r-1,1)$-sparse.
A $(0,1)$-sparse set of qubits in $M_0$ is an empty set of qubits.  
\item
A set of locations in a $r$-rectangle is $(r,1)$-sparse when there is at most $1$ $(r-1)$-rectangle
 such that the set is not $(r-1,1)$-sparse in that $(r-1)$-rectangle.
A fault-path in $M_r$ is $(r,1)$-sparse if in
 each $r$-rectangle, the set of faulty locations is $(r,1)$-sparse.
 \item
A computation code $C$ has spread s if one fault which occurs in a
particular 1-rectangle affects at most s
qubits in each 1-block at the end of that 1-rectangle,
i.e. causes at most $s$ errors in each 1-block.
\item
Let $A_C$ be the number of locations in a $1$-rectangle for a given code $C$.}
\end{itemize}

We state the basic lemma about properties of computation codes which was proved in Ref. \cite{AB-ftsiam} (with
a correction).

\begin{lem}[{\sf A}: Lemma 8 in \cite{AB-ftsiam} with a correction] Let $C$ be a
computation code that can correct 2 errors and has spread $s=1$.
Consider a computation $M_r$ subjected to a $(r,1)$-sparse
fault-path. At the end of each $r$-working period the set of
errors is $(r,1)$-sparse. \label{codelem}
\end{lem}

Thus for simplicity we will be using a quantum computation code that encodes one qubit and
can correct 
two errors and has spread $s=1$.  
We denote the entire unitary evolution of $M_r$ including the bath as $Q^r$.
We may write $Q^r=Q_G^r+Q_B^r$ where $Q_G^r$ is a sum over {\em good} $(r,1)$-sparse fault-path operators and
$Q_B^r$ contains the {\em bad} non-sparse terms. A fault-path operator $E_{\texttt{SB}}$ that is $(r,1)$-sparse
is a sequence of free evolutions $U_0[i]$ for all locations except that in every
$r$-rectangle there is a $(r,1)$-sparse set of locations where a fault operator $E[i]$ occurs. 

\begin{defi}[Operators in the Interaction Picture]
Let $U_0(t_2,t_1)=U_{\texttt{S}}(t_2,t_1) \otimes U_{\texttt{B}}(t_2,t_1)$ be the free uncoupled
evolution of system and bath in the time-interval $[t_1,t_2]$. We define a fault-operator $E(t_2,t_1)$ in the interaction picture as
\beq
E(t_2,t_1)=U_0(t_2,t_1) E\,  U_0^{\dagger}(t_2,t_1).
\eeq
The interpretation is that $E(t_2,t_1)$ is the spread of a fault $E$ that occurs at $t_1$ due to
the subsequent free evolution.
\end{defi}

Then it is simple to see the following:

\begin{propo}[Error Spread in the Interaction Picture]
Consider a quantum circuit $M$. Let $U_0(t_F,t_I)$ be the uncoupled evolution for $M$.
Faults occur at a set of `time-resolved' locations
\[\mathcal{T}=((i_1,t_1),(i_2,t_2),\ldots,(i_k,t_k))\]     
where $i_1,\ldots,i_k$ is the set of distinct locations of the faults and
$t_1,\ldots,t_k$ label the specific times that the faults occur at
the locations. Let $E_{\texttt{SB}}(\mathcal{T})$ be a particular fault-path
operator in which at every faulty location $(i,t) \in \mathcal{T}$ we replace $U_0[i]$ by a fault-operator
$E[i]$. We have
\beq
E_{\texttt{SB}}(\mathcal{T}) U_0^{\dagger}(t_F,t_I)=E[i_k](t_F,t_k) \ldots E[i_1](t_F,t_1).
\eeq
We note that the system part of $E_{\texttt{SB}} U_0^{\dagger}$ is
${\bf I}$ everywhere except for the qubits that are in the causal cone of the faulty locations, i.e.
the qubits to which the errors potentially have spread.
\label{lemspread}
\end{propo}

\begin{proof}
This can be shown by inserting ${\bf I}=U_0^{\dagger}(t_F,t_i)U_0(t_F,t_i)$ in the appropriate places and then
using the definition of fault operators in the interaction picture.
\end{proof}

Now we include error-correction and differentiate between the
ancilla systems \texttt{A} used for error-correction which may
contain noise and the registers \texttt{R} in which the errors
remain sparse. Note that all these ancillas are in principle
discarded after being used, but we may as well leave them around.
Let $K |_C$ be the restriction of the operator $K$ to vectors in
the code-space of $C$, i.e. $K|_C=K\, {\bf P}_C$ where ${\bf P}_C$
is the projector on the codespace.


Let us consider a fault-path operator $E_{\texttt{SB}}$ representing a single fault $E$ at time $t$ on some block that
is subsequently corrected by an errorfree error-correcting procedure. Let $\ket{{\rm IN}}$ be the initial
state of the computer, bath and ancillas and $U_0(t_F,t_I)$ be the perfect evolution.
We have
\beq
E_{\texttt{SB}} \ket{{\rm IN}}=E_{\texttt{SB}} U_0^{\dagger} U_0 \ket{{\rm IN}}=E_{\texttt{SB}} U_0^{\dagger}
\ket{\psi_C(t_F)}.
\eeq
where $\ket{\psi_C(t_F)}$ is the final perfect state of the computer, prior to decoding and therefore
in the code-space. $E_{\texttt{SB}}$ is the sequence $U_0(t_F,t) E\, U_0(t,t_0)$ where $U_0(t_F,t)$ includes
the error correction operation. In other words, in the interaction picture, we can write
\beq
E_{\texttt{SB}} \ket{{\rm IN}}=E(t_F,t) \ket{\psi_C(t_F)},
\label{exam}
\eeq
The error-correcting conditions (see \cite{book-nielsen&chuang}, par. 10.3) imply that when acting
on the code space {\em and} an ancilla state set to $\ket{00 \ldots 0}$ the operator $E(t_F,t)$
will be $E(t_F,t)={\bf I}|_C \otimes ({\rm Junk})_{\texttt{AB}}$ where ${\rm Junk}$ is some arbitrary operator on the
ancilla (that receives the error syndrome in the error-correcting procedure) and bath. In Eq. (\ref{exam}) the
final errorfree state has all ancillas set to $\ket{00 \ldots 0}$ and the system state is in the code-space and thus
the error acts as ${\bf I}$ on the system.

Similarly, let $E_{\texttt{SB}}$ contain   
two faults at times $t_1 < t_2$ that have not spread (say) and are then corrected
by a perfect error-correcting procedure. We have
\beq
E_{\texttt{SB}} \ket{{\rm IN}}=E_2(t_F,t_2) E_1(t_F,t_1) \ket{\psi_C(t_F)}.
\eeq
Let us assume, for example, that $E_1$ occurs prior to error-correction and
$E_2$ occurs during error-correction.
Then due to the error correction $E_1(t_F,t_1)$ acts as ${\bf I}$ on the code space when the ancilla
used for error-correction is set to $\ket{00 \ldots 0}$ and acts as ${\rm Junk}$ on this ancilla and the bath.
The error $E_2$ will not be corrected and may still be present (but will not have
spread to more qubits in the block due to the spread properties of the code that is used) after error-
correction. Thus in total we can write for this process that $E_1(t_F,t_1)$ acts as ${\bf I}$ on the
code space, whereas $E_2(t_F,t_2)$ 
is an operator that acts on the code space as at most one error per block.

Alternatively, both faults could occur prior to error-correcting so they can both be corrected by our code.
This implies that both $E_1(t_F,t_1)$ and $E_2(t_F,t_2)$ act as ${\bf I}$ on the code-space. Note that after the first
fault the ancilla will be partially filled (i.e. not be $\ket{00 \ldots 0}$) but since the code can correct
two errors there is still space to put the second error syndrome in. However a third operator $E_3(t_F,t_3)$ would no
longer act as ${\bf I}$ on the code-space since the code cannot correct three errors.

In other words, with these examples we can see how Lemma \ref{codelem} can be translated in
terms of the sparseness of the errors in the interaction picture, i.e. the sparseness of
places where they act as non-identity on the final encoded state of the register qubits.
In the next lemma we need to consider the effect of such sparse fault-path operators
$E_{\texttt{SB}}$ on the final state of the computer. This is the state
of the computer obtained after fault-tolerant decoding which is as follows.
The fault-tolerant decoding procedure for a single level of encoding takes
a codeword $\ket{c}$ and `copies' (by doing CNOT gates) the codeword $m$ times. Then on each `copy'
we determine what state it encodes and then we take the majority of the $m$ answers.
This procedure is done recursively when more levels of encoding are used.

In the fault-tolerant decoding procedure faults can occur on the codewords, i.e. as incoming faults,
during the copying procedures and during the determination of what is encoded by the codeword. The last
procedure will usually be a conversion from a quantum state to a classical bit string since
this will be the most efficient. This implies that the step of taking the recursive majority of
these bits is basically errorfree since it only involves classical data. In the next Lemma we model
this by coherent quantum operations that output superpositions of decoded bit strings followed
by an error-free measurement that takes the recursive majority of these bits.

\begin{lem}[{\sf B}: Sparse faults give almost correct answers]
Let $Q^r=Q_G^r+Q_B^r$ the unitary evolution of $M_r$ and let $||Q_B^r|| \leq \epsilon < 1/2$.
Let $\mathbb{P}_0(i)$ be the output probability distribution under
measurement of some set of qubits of the error-free original computation $M_0$.
Let $\mathbb{P}(i)$ be the simulated output distribution of the encoded computation $M_r$
with evolution $Q^r$. We have
\beq
||\mathbb{P}_0-\mathbb{P}||_1 \leq \sqrt{2\epsilon}+16 \epsilon.
\eeq
\label{dev}
\end{lem}

\begin{proof} The initial state of the computer is $\ket{{\rm IN}}_{\texttt{RAB}}=\ket{00 \ldots 0}_{\texttt{RA}} \otimes \ket{{\rm IN}_{\texttt{B}}}$
for some arbitrary state $\ket{{\rm IN}_{\texttt{B}}}$.
Let $U_0^r$ be the error-free evolution of $M_r$ including the final decoding operation.
Thus let $U_0^r \ket{{\rm IN}}_{\texttt{RAB}}=\ket{{\rm OUT}_0}_{\texttt{R}}\otimes \ket{{\rm REST}}_{\texttt{AB}}$.
Let $Q^r \ket{{\rm IN}}_{\texttt{RAB}}=\ket{{\rm OUT}}_{\texttt{RAB}}$ and
$Q^r_{B/G}\ket{{\rm IN}}_{\texttt{RAB}}=\ket{{\rm OUT}_{B/G}}_{\texttt{RAB}}$. We will drop the label $\texttt{RAB}$ from now on.
The norm of $\ket{{\rm OUT}_G}$ will be denoted as $||{\rm OUT}_G||$.
We have
\beq
1=||Q^r \ket{{\rm IN}}||\leq ||Q_G^r \ket{{\rm IN}}||+||Q_B^r\ket{{\rm IN}}||,
\eeq
so that $||{\rm OUT}_G|| \geq 1-||Q_B^r\ket{{\rm IN}}||\geq 1-\epsilon$. On the other hand $||Q_G^r||=||Q^r-Q_B^r|| \leq 1+\epsilon$.

Let $G$ be the set of $(r,1)$-sparse fault-paths. We have
$Q^r_G=\sum_{\mathcal{T} \in G} E_{\texttt{SB}}(\mathcal{T})$
where $E_{\texttt{SB}}(\mathcal{T})$ is the fault-path operator of
a $(r,1)$-sparse fault-path labelled by location and time index
set $\mathcal{T}$.
 We can write
\beq \ket{{\rm OUT}_G}=\sum_{\mathcal{T} \in G}
E_{\texttt{SB}}(\mathcal{T}) {U_{0}^r}^{\dagger} \ket{{\rm
OUT}_0}_{\texttt{R}}\otimes \ket{{\rm REST}}_{\texttt{AB}}. \eeq
By the arguments above and the fundamental Lemma \ref{codelem} we
know that $E_{\texttt{SB}}(\mathcal{T}) U_0^{\dagger}$ is ${\bf
I}$ everywhere except on a $(r,1)$-sparse set of qubits. Let $w$
be the number of output qubits of $M_0$. The ideal state
$\ket{{\rm OUT}_0}_{\texttt{R}}$ has the property that all qubits
in an $r$-block have the same value in the computational basis,
i.e. \beq \ket{{\rm OUT}_0}_{\texttt{R}}=\sum_{i_1,\ldots, i_w}
\alpha_{i_1 \ldots i_w} \ket{i_1}^{\otimes m^r}\ldots
\ket{i_w}^{\otimes m^r}, \eeq where $m$ is the number of qubits in
a $1$-block. The final step of the computation is a measurement of
all output qubits that takes the recursive majority on the block
to get the final output string $i$ of length $w$ with probability
$\mathbb{P}^{tot}(i)$. We model this measurement using POVM
elements $E_k$, -- $\sum_k E_k={\bf I}$.  Since not all these $w$
output bits may be relevant output bits of $M_0$, we may use the
fact that trace-distance is non-increasing over tracing
\cite{book-nielsen&chuang} so that \beq
||\mathbb{P}_0-\mathbb{P}||_1 \leq
||\mathbb{P}^{tot}_0-\mathbb{P}^{tot}||_1, \eeq where
$\mathbb{P}^{tot}(k)={\rm Tr} E_k \ket{{\rm OUT}}\bra{{\rm
OUT}}_{\texttt{RAB}}$ and $\mathbb{P}^{tot}_0(k)={\rm Tr} E_k
\ket{{\rm OUT}_0}\bra{{\rm OUT}_0}_{\texttt{R}}$. Let us also
define $\mathbb{P}^{tot}_G$, the distribution of outcomes if the
state of the computer would be the normalized state $\ket{{\rm
OUT}_G^N} \equiv \ket{{\rm OUT}_G}/|| {\rm OUT}_G||$. The triangle
inequality and the properties of the trace-norm imply that
\bea
||\mathbb{P}^{tot}-\mathbb{P}^{tot}_0||_1 \leq
||\mathbb{P}^{tot}-\mathbb{P}^{tot}_G||_1 +||\mathbb{P}^{tot}_G-
\mathbb{P}^{tot}_0||_1 \leq \nonumber \\
||\,\ket{{\rm OUT}}\bra{{\rm
OUT}}-\ket{{\rm OUT}_G^N}\bra{{\rm OUT}_G^N}\,||_1+
||\mathbb{P}^{tot}_G-\mathbb{P}^{tot}_0||_1. \label{bind}
\eea
Here the first term can be bounded, using the relation of the
trace norm to the fidelity $F(\psi,\phi)=|\bra{\psi}\phi\rangle|$
\cite{book-nielsen&chuang},
 as
\bea
||\,\ket{{\rm OUT}}\bra{{\rm OUT}}-\ket{{\rm OUT}_G^N}\bra{{\rm OUT}_G^N}\,||_1 \leq
\sqrt{1-F\left({\rm OUT},{\rm OUT}_G^N\right)^2} \leq \nonumber \\
\sqrt{1-||{\rm OUT}_G||^2}\leq \sqrt{2 \epsilon-\epsilon^2}.
\eea
Now consider the second tracenorm on the r.h.s. of Eq. (\ref{bind}). We note that
all states that are linear combinations of $(r,1)$-sparse error sets applied to the state
$\ket{k_1}^{\otimes {m}^r}\ldots \ket{k_w}^{\otimes {m}^r}$ will give rise to the
measurement outcome $k$ since we are taking majorities. We can model $E_k=P_k$ where
$P_k$ is the projector onto the space of computational basis states that give rise to the majority output
string $k$. Thus we have
\beq
P_k \ket{{\rm OUT}_G}=\alpha_{k_1 \ldots k_w}\sum_{\mathcal{T} \in G} E_{\texttt{SB}}(\mathcal{T}) {U_{0}^r}^{\dagger} \ket{k_1}^{\otimes {m}^r}\ldots \ket{k_w}^{\otimes {m}^r} \otimes \ket{{\rm REST}}_{\texttt{AB}}.
\eeq
which can be written as $\alpha_{k_1 \ldots k_w} Q_G^r \ket{\psi_k}$ for some normalized state $\ket{\psi_k}_{\texttt{RAB}}$.
This implies that the second term in Eq. (\ref{bind}) can be bounded as
\bea
||\mathbb{P}^{tot}_G-\mathbb{P}_0^{tot}||_1= \sum_k |\alpha_{k_1 \ldots k_w}|^2 \left|\,\frac{||Q_G^r \ket{\psi_k}||^2}{||{\rm OUT}_G||^2}-1\right| \leq
\nonumber \\
\sum_k  |\alpha_{k_1 \ldots k_w}|^2
\max_k \left|\,\frac{||Q_G^r \ket{\psi_k}||^2}{||{\rm OUT}_G||^2}-1\right| \leq \frac{4 \epsilon}{(1-\epsilon)^2},
\eea
using the bounding inequalities of $||Q_G^r||$ and $||{\rm OUT}_G||$. All bounds put together, using $\epsilon < 1/2$,
give the result, Eq. (\ref{dev}).
\end{proof}

\subsection{Step {\sf C}: non-sparse fault-paths have small norm}
\label{secC}

Consider the evolution $Q^r$ which can be viewed as a sequence of unitary evolutions, one for each $r$-rectangle, since qubits in different rectangles do not interact.
The number of locations in $M_0$ is $N$. The computation $Q^r$ is \emph{bad} when at least one 
$r$-rectangle is bad, or using Lemma \ref{blemma1}
\beq
||Q_B^r|| \leq N ||R_B^r||\, ||R_G^r||^{N-1},
\eeq
where $R_B^r$ and $R_G^r$ are the good and bad parts of the
unitary evolution $R^r$ for some $r$-rectangle. The unitarity of $R^r$ implies that
we can bound $||R_G^r|| \leq 1+||R_B^r||$. In each rectangle we can view the entire evolution as a sequence of unitary evolutions, one for each $(r-1)$-rectangle.
Note that we are again using the fact that non-interacting qubits have non-interacting baths. A $r$-rectangle
is bad when there are at least two 
$(r-1)$-rectangles which contain sets of faulty locations which
are not $(r-1,1)$ sparse. This implies, using Lemma \ref{blemma1} again, that
\beq
||R_B^r|| \leq {A_C \choose 2}||R_B^{r-1}||^2 ||R_G^{r-1}||^{A_C-2},
\label{recur}
\eeq
where we can use that $||R_G^{r-1}||\leq 1+||R_B^{r-1}||$.
 When $r=1$, $R_G^1$ is a unitary operation and thus
$||R_G^1||=1$. This recurrence in $r$ is identical to the one in Lemma 11 in Ref. \cite{AB-ftsiam} and
thus the solution and results are the same if we replace $\eta$ in Ref. \cite{AB-ftsiam} by $\lambda_0 t_0$.
Thus the critical error threshold value is
\beq
(\lambda_0 t_0)_c=\frac{1}{{\bf e} A_C (A_C-1)}.
\label{threshval}
\eeq

Here we can observe a difference with the simplest error model with error probability $p$ for which
the critical value is $p_c=\frac{1}{{A_C \choose 2}}$ \cite{AB-ftsiam}.
The dimensionless quantity $\lambda_0 t_0$ plays the role of an {\em amplitude},  
see Sec. \ref{strength}, which implies that this threshold value may be more 
stringent than in the
simple probabilistic error model (see also the critique by Alicki \cite{alicki:crit} on
our results). However, we believe that this analysis is
too course to really give information about the value of the threshold. 
The fact is that in practice, baths do not have
infinite memory times since they are coupled to many other degrees of freedom.
Representing the coupling between bath and system as a pure coherent evolution was
needed in this analysis to deal with the non-Markovian dynamics; 
however, we do not expect this formal procedure to give rise to an 
\textit{optimal} error threshold.

The idea of the remaining derivation given in Ref. \cite{AB-ftsiam} is to show that
when $\lambda_0 t_0 < (\lambda_0 t_0)_c$ for large enough concatenation level r, $||Q^r_B||\leq \epsilon$
for arbitrary small $\epsilon$.  Lemma \ref{dev} then tells how much our
quantum computation errs from the error-free computation. Summarizing we get the following, as in
Ref.~\cite{AB-ftsiam}:

\begin{theo}[Threshold Theorem for Local Non-Markovian Noise]
Let $N$ be the number of locations of an errorfree quantum computation $M$
that outputs samples from a probability distribution $\mathbb{P}$.
There exists a quantum circuit $M'$ subjected to noise according to the Hamiltonian $H_{\texttt{SB}}$ and
bath Hamiltonian $H_{\texttt{B}}$ that obeys the ``noninteracting qubits have noninteracting baths''
assumption which outputs the probability distribution $\mathbb{P'}$ such that
\beq
||\mathbb{P'}-\mathbb{P}||_1 \leq \epsilon,
\eeq
when $\Delta_{\texttt{SB}} t_0 < \frac{1}{{\bf e} A_C (A_C-1)}$ and $A_C$ is the number of
locations in a $1$-rectangle of a (computation) code C that can correct two 
errors and has spread $s=1$.
The number of locations in $M'$ 
is $N {\rm polylog} \frac{N}{\sqrt{\epsilon}}$.
\label{thresh1}
\end{theo}

\begin{proof}
This follows from the basic results in Ref. \cite{AB-ftsiam} and Lemma \ref{dev} and step ${\sf C}$ above.
It was proved in Ref. \cite{AB-ftsiam} for $\lambda_0 t_0$ below the threshold
$(\lambda_0 t_0)_c$ in Eq. (\ref{threshval}) when the concatenation level
$r=c_1 \log (\log \frac{N}{\epsilon'}+c_2)+c_3$ for constants $c_1$, $c_2$ and $c_3$ we have $||Q_B^r|| \leq \epsilon'$.
So we choose $M'=M_r$, the computation at this concatenation level $r$ which
implies that $||\mathbb{P'}-\mathbb{P}||_1 \leq \sqrt{2 \epsilon'}+16 \epsilon' \equiv \epsilon$.
The number of resources (time and space, related to $A_C$) in $M'$ scales exponentially, i.e. the number of
locations in $M_r$ is $N A_C^r$. With the dependence of $r$ on $N$ and $\epsilon'$ this implies the
polylogarithmic overhead in terms of $N$ and $\sqrt{\epsilon}$.

\end{proof}


\section{Extension to Decoherence Models with Clustered Qubits}
\label{bathext}

In the most general noise model we start with a Hamiltonian description of
system and bath. We will assume that such Hamiltonians are 1-\emph{systemlocal}, that is,
the interaction Hamiltonian between system and bath is a sum of terms each of which couples a
{\em single} qubit to some part of the bath. This covers many
interaction Hamiltonians in systems that are being considered for quantum computation (see footnote \cite{fn4}).

We have seen that basically the only place where the noise model enters
the derivation of fault-tolerance is in section \ref{secC}, i.e. the derivation
that the total amplitude/probability/norm for non-sparse fault-paths at concatenation level $r$ goes (doubly exponentially
fast in $r$) to zero when the initial error strength is below the threshold. Locality
of the interaction Hamiltonian is an important (and necessary) ingredient in the
derivation of fault-tolerance since it implies --without any further assumptions on the structure of the bath
or the (Non)-Markovian character of the system-- that fault-path operators with $k$ faults have a
norm bounded by $(2 \lambda_0 t_0)^k$ (see Appendix \ref{appendixf}).
This bound is not strong enough by itself to derive that $||Q_B^r||$ becomes arbitrarily 
small for sufficiently small $\lambda_0 t_0$. We find that there are
technical and potentially fundamental problems in the derivation of step {\sf C}, for
the most general local Hamiltonian model both {\em in the Markovian case} as well
as in the non-Markovian case. The problems are due to the fact that all qubits of
the computer potentially couple at a given time to the same bath which was prevented in the derivation
of Theorem 1 by assuming that ``noninteracting qubits have noninteracting baths". The problem is basically
due to the fact that the unitary evolution
of a working period cannot be written as a product of unitary evolutions for each rectangle in the working
period since different rectangles may share their bath.

We thus need to consider restricted models that are still physically very relevant:

\subsection{Clustered Qubits at encoding level $r=1$}

\begin{figure}[htb]
\begin{center}
\epsfxsize=15cm \epsffile{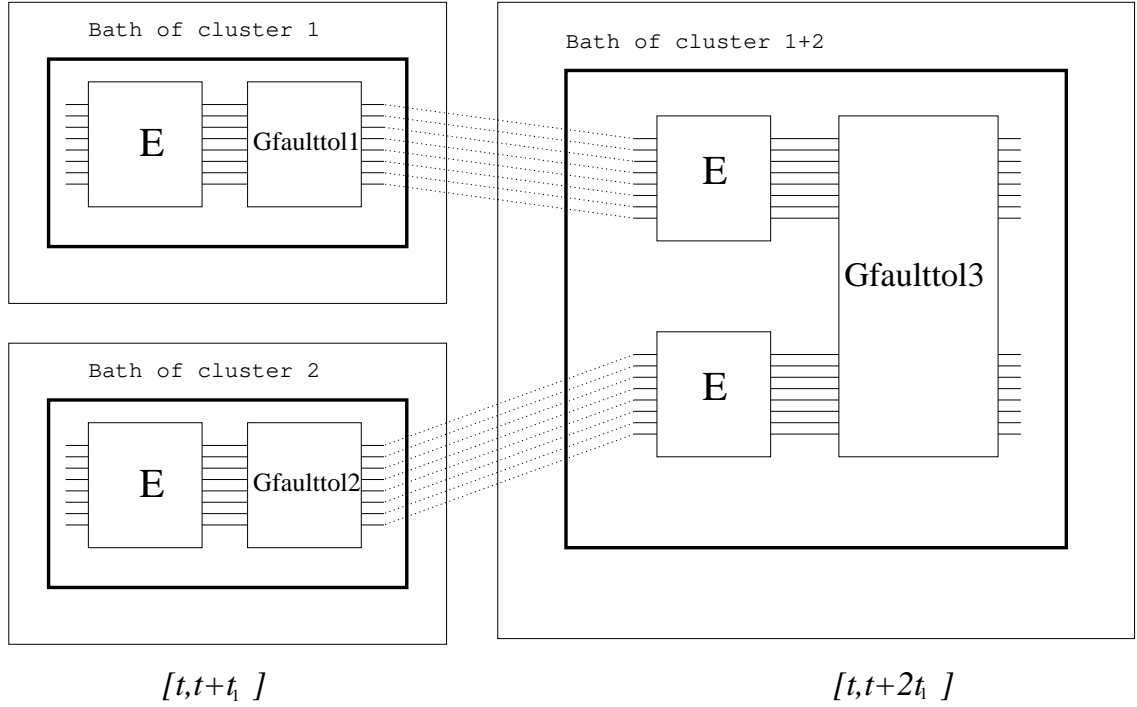} \caption{Schematic
representation of decoherence model where clusters of qubits can
share a common bath. Logical qubits 1 and 2 are encoded once in a
block of qubits. In the original circuit these qubits first
undergo single qubit gates $G_1$ and $G_2$ and then interact in
$G_3$. In encoded form this implies three 1-rectangles that each
take some time $t_1$; these are denoted by the boxes with fat
lines in the figure. Each 1-rectangle or cluster has its own bath.
These baths may change over time, that is, the bath of cluster 1
may evolve or change and not be the same as the environment that
this block of qubits sees later.} \label{figschem2}
\end{center}
\end{figure}

We can generalize the model in Sec. \ref{decohmodel}, i.e. noninteracting qubits have noninteracting baths,
to one in which a cluster of qubits can share a bath. The model is depicted in Fig. \ref{figschem2}.
We will assume that qubits that are contained in a $1$-rectangle of $M_r$ may share a common bath
whereas qubits in different $1$-rectangles do not share a bath. We imagine that baths are attached to physical locations,
so that the interaction regions of different $r$-rectangles are physically separate. This means that from one $1$-working
period to the next one, qubits have to be moved around, i.e. qubits that participate in one $1$-rectangle have to
be brought together.

Let us for the moment neglect the machinery that is necessary to move qubits around. Then we can observe that
the entire computation $M_r$ can be viewed as a sequence of unitary gates each involving a single
 $1$-rectangle. In the $1$-rectangle we cannot decompose the evolution as a sequence of unitary transformations for
 each location. But at this lowest level $r=1$ it is simple to derive a bound on the \emph{bad} part $R_B^1$ of
 the unitary operation $R^1$. Given this bound we can insert it in the previous recurrence of Eq. (\ref{recur})
and determine a threshold which is the same as before. Here is the bound on $R^1_B$:

\begin{lem}
Let $R^1=R^1_B+R^1_G$ be the unitary transformation of a $1$-rectangle where $R^1_B$ is a sum of
non-sparse fault-path operators, i.e. each such operator contains at least two locations with faults.
Then
\beq
||R^1_B|| \leq 2(A_C t_0 \lambda_0)^2.
\eeq
and $||R^1_G|| \leq 1+||R^1_B||$.
\label{enc1}
\end{lem}

\begin{proof}
We do a Trotter expansion for $R^1$ as in Lemma \ref{fpbound} and obtain a tree with infinite depth.
We combine branches of the tree in the following way: 1. after a location has become faulty we append
the full unitary for the remaining time of the location and 2. if two faults have occurred at two different
locations we do no longer branch the tree and just append the entire remaining unitary transformation to
that branch. In this way the norm of every time-resolved branch with at least two faults $E_{2^+}(\mathcal{T})$
is bounded by $||E_{2^+}(\mathcal{T})|| \leq  ( 2\lambda_0 t_0/n)^2$. There are ${A_C t_0 n/t \choose 2}$ such
branches and thus
\beq
||R_B^1|| \leq ( 2\lambda_0 t/n)^2 {A_C t_0 n/t \choose 2} \leq 2(A_C t_0 \lambda_0)^2.
\eeq
\end{proof}


A physical example of this decoherence model is the proposal for scalable ion-trap computation \cite{wineland+-iontraps}.
A few qubits are stored in an ion-trap where they may share a common bath. The states of
qubits can be moved around to let them interact. A small cluster of ion-traps may be used to
carry out the fault-tolerant circuits and error-correcting at level $r=1$ of encoding.

The issue of moving qubit states around is not entirely trivial and will
be addressed in detail in a future paper \cite{STD-local}.

\section{Overview}
\label{soa}

\begin{table}[h]
\begin{tabular}{l|l|c|c|c}
\multicolumn{4}{r}{{\sf Spatial Correlations}} \\
\multicolumn{5}{r}{} \\
& & Single Location & Cluster Location Baths & Arbitrary   \\
& & Baths & at $r=1$  & Baths \\
\hline
\multirow{3}{1in}{{\sf Temporal Correlations}} & Markovian within & $\checkmark$ & $\checkmark$ & ? \\
& gate-time $t_0$ & & & \\
\cline{2-5}
& Non-Markovian with & $\checkmark$ & $\checkmark$  & ? \\
& finite memory time $\tau > t_0$  & &  & \\
\cline{2-5}
& Non-Markovian with  & $\checkmark$ & $\checkmark$ &  ?\\
& unlimited memory time & & & \\
\cline{2-5}
\end{tabular}
\caption{A $\checkmark$ indicates that a fault-tolerance result
exists whereas a question mark ? indicates that it is not known to
exist so far (neither has it been disproved). The results for
non-Markovian baths assume a 1-systemlocal interaction Hamiltonian
that can be bounded in norm. They also assume that we can do
two-qubit gates between any two qubits in the circuit (that is, we
do not take physical locality constraints into account). The
assumptions on the structure of the system-bath interaction and
the bath Hamiltonian are given by the three columns. Single
Location Baths implies that the interaction and the baths are
constrained so that for each elementary time-interval
(clock-cycle) $[t,t+t_0]$ the following condition is obeyed:
qubits that do not interact can only interact with baths which do
not interact, see Fig. \ref{figschem}(b). Note that the particular
baths with which the qubits interact may change over time. Cluster
Location Baths is the extension of this model covered in Section
\ref{bathext} where a cluster of qubits can share the same bath,
see Fig. \ref{figschem2}. In the last column there is no
constraint on the bath.}
\end{table}

We would like to summarize the known results, including the ones in this paper, on threshold results
for different decoherence models. The simplest model is one in which we assume that each location undergoes an error with
probability $p$ and undergoes no error with probability $1-p$. This is a specific example of a Markovian model in
which in every location has its own separate environment, i.e we have ``Single Location Baths'', see the upper left entry
in the Table 1. Generalizations of this model exist \cite{AB-ftsiam,klz-faulttol};
in these models a superoperator ${\cal S}(\rho)={\cal S}_0(\rho)+{\cal E}(\rho)$ where ${\cal S}_0$ corresponds to the error-free evolution
and ${\cal E}$ to the erroneous part, is associated with each location. Again this corresponds to the upper left entry in the table.
This model has been generalized to allow for more general correlations in space and time in
the following manner. In Ref. \cite{AB-ftsiam} fault-tolerance was derived in a
model where it is assumed that the probability for a fault-path with $k$ faults is bounded by
$ C p^k (1-p)^{N-k}$ where $N$ is the total number of locations in the circuit (note the difference with
Eq. (\ref{fpboundeq}) in Appendix \ref{appendixf}). Similarly, in
Ref. \cite{klz-faulttol} fault-tolerance was derived under the assumption that a fault-path with {\em at least} $k$ faults has probability bounded by $C p^k$ for some
constant $C$. Let us call these conditions the exponential
decay conditions.
Note in the Table that it is not known whether
one can derive fault-tolerance for a entirely Markovian model but {\em with} extended
spatial correlations between the baths, i.e. for every clock-cycle we have a superoperator that
acts on all qubits of the system; the point is that it is not clear whether such a
superoperator would obey some sort of exponential decay conditions.


\section{Measures of Coupling Strength and Decoherence}
\label{strength}

In our analysis the role of error amplitude is played by the dimensionless number $\lambda_0 t_0$
which captures the relative strength of the interaction Hamiltonian as compared to the system Hamiltonian.
It is this quantity, $\lambda_0 t_0$, that should be $O(10^{-4})$ as was determined for some
codes. In a purely Markovian analysis we
 typically replace $\lambda_0$ by an inverse $T_2$ or $T_1$ time and this may
 give a more optimistic idea of the regime of fault-tolerance.
Let us consider a few examples of decoherence mechanisms and see
how sensible it is to use $\Delta_{\texttt{SB}}$ as a bound for
decoherence. A good example of a non-Markovian decoherence
mechanism is a small finite dimensional environment localized in
space, for example a set of spins nearby the system of interest.
An example is the decoherence in NMR due to interactions with
nuclear spins in the same molecule. In NMR the nuclear exchange
coupling between spins $a$ and $b$ is given \beq H_{int}=J_{ab}
\;\vec{I}_a \cdot \vec{I}_b. \eeq If the J-coupling is treated as
a source of decoherence as compared to the Zeeman-splitting
$\omega_0$ for an individual spin, then $J/\omega_0$ can be $\sim
10^{-6}$ (see footnote \cite{fn5}).


For some physical systems a source of decoherence is a bath of spins, each of which couples to a single qubit.
An example is the electron spin qubit in a single quantum dot which couples via the hyperfine coupling to
a large set of nuclear spins in the semiconductor \cite{KLG-esd}. The interaction Hamiltonian is as follows
\beq
H_{int}=\sum_{i=1}^N a(i) \; \vec{\sigma}\cdot {\vec I}[i],
\eeq
where $a(i)=A v_0 |\psi_s(i)|^2$ and $A$ is the hyperfine coupling constant, $v_0$ the volume of
the crystal cell, and $|\psi_s(i)|^2$ is the probability of the electron to be
at the position of nuclear spin $i$. If we bound $\sum_i |\psi_s(i)|^2\le 1$ we have that
$||H_{int}|| \leq C A v_0$ where $C$ is a small constant (of order 1). This may give a somewhat weak upper bound
on the decoherence, since we are basically adding the effects of each nuclear spin separately.

A third type of decoherence mechanism exists which is essentially troublesome in our analysis. This is
the example of a single qubit, or spin, coupled to a bosonic bath. The interaction Hamiltonian is that of the
spin-boson model \cite{book-weiss}
\beq
H_{int}=\sigma_z \otimes \sum_{i=1}^N (c_i a_i+c_i^* a_i^{\dagger}),
\eeq
where $i$ labels the $i$th bosonic mode characterized by frequency $\omega_i$. The $i$th bosonic mode has
Hamiltonian $H_{\texttt{B}_i}=\omega_i (a_i^{\dagger}a_i+\frac{1}{2})$.
In order to represent a continuous bath spectrum, one
lets $N$ go to infinity. In that limit the coupling constants $c_i$ are determined by the
spectral density $J(\omega)=\sum_i |c_i|^2 \delta(\omega-\omega_i)$. The spectral density can have
various forms, matching the phenomenology of the particular physical system, an example is the
Ohmic form in which $J(\omega)=\alpha \omega e^{-\omega/\omega_c}$ where $\alpha$ is a weak coupling
constant that has physical relevance and $\omega_c$ is a cutoff frequency that is also determined by
the physics. It is clear that $||H_{int}||$ has no physical meaning since it is infinite, the reason being
that there are infinitely excited bath states with infinitely high energy. We can determine an energy-dependent
upper bound on this norm; using properties of the norm, we can estimate
\bea
||H_{int}\ket{\psi}_{\texttt{SB}}||= ||\sum_i c_i a_i+c_i^* a_i^{\dagger}) \ket{\psi}_{\texttt{SB}}|| \leq
\sum_i |c_i|(||a_i \ket{\psi}_{\texttt{SB}}||+||a_i^{\dagger} \ket{\psi}_{\texttt{SB}}||) \nonumber \\
\leq  \sum_i \frac{|c_i|}{\sqrt{\omega_i}} \sqrt{4 \bra{\psi} H_{\texttt{B}_i} \ket{\psi}_{\texttt{SB}}}.
\label{boundbos}
\eea
Using the Schwartz inequality we get
\beq
||H_{int}\ket{\psi}_{\texttt{SB}}|| \leq 2 \sqrt{\sum_i \frac{|c_i|^2}{\omega_i} \bra{\psi} H_{\texttt{B}} \ket{\psi}_{\texttt{SB}}}
=2 \sqrt{\langle H_{\texttt{B}} \rangle_{\psi_{\texttt{SB}}} \int_0^{\infty} d\omega \frac{J(\omega)}{\omega} }.
\label{hb-bound}
\eeq
for some state of system and bath $\ket{\psi}_{\texttt{SB}}$ where the bath Hamiltonian $H_{\texttt{B}}=\sum_i H_{\texttt{B}_i}$.
The idea is
that for the physically relevant states of the bath $\langle H_{\texttt{B}} \rangle_{\psi}$ is bounded.
The problem
remains that this bound will in general be too poor to be physically relevant, since this energy bound may be
quite large. Also, for ohmic coupling (for example) we have
the integral $\int_0^{\infty} d\omega \frac{J(\omega)}{\omega}=\alpha \, \omega_c$, i.e. linear in $\omega_c$.
The cutoff $\omega_c$ may be quite large and it is more typical to see decoherence rates depend on
$\log \omega_c$ as in the non-Markovian analysis of Ref. \cite{LD-spinbos} for example.

\subsection{Cooling assumption} 
Some progress can be made in finding good bounds for $||H_{int}||$
in the case of a bosonic environment if additional assumptions
about its state can be made. What is troublesome about the
potential nonequilibrium state of the bath is that expectation values 
such as ${\rm Tr}\, a_i a_j \ket{\psi}\bra{\psi}_{\texttt{SB}}$
and ${\rm Tr}\, a_i^2 \ket{\psi}\bra{\psi}_\texttt{SB}$ may not be
zero since the bath state may not be 
diagonal in the energy or boson
number basis. On the other hand, interaction with other
environments, for example by means of cooling, will dephase the
state of the bath (due to energy exchange) and drive it to a state
that is diagonal in the energy eigenbasis. Under that assumption
only the terms ${\rm Tr}\, a_i^{\dagger}a_i
\ket{\psi}\bra{\psi}_{\texttt{SB}}$ and ${\rm Tr}\,a_i
a_i^{\dagger} \ket{\psi}\bra{\psi}_{\texttt{SB}}$ are nonzero. In
that scenario, Eq. (\ref{boundbos}) simplifies to \beq
||H_{int}\ket{\psi}_{\texttt{SB}}|| \approx \sqrt{\sum_i |c_i|^2
\bra{\psi} a_i^{\dagger}a_i+{\bf I}/2) \ket{\psi}_{\texttt{SB}}}.
\eeq Still, $||H_{int}\ket{\psi}_{\texttt{SB}}||$ can be very
large if some modes of the environment are highly excited,
$n_i=a_i^{\dagger}a_i \gg 1$. However, in a realistic setting,
this will be prevented by cooling the bath, i.e.\ by constantly
removing energy from it. Without making a Markov approximation, we
can thus assume that the occupation numbers $n_i$ of the bath are
upper-bounded by those of a thermal distribution with an effective
maximal temperature $T_{\rm eff}$. This gives
\begin{equation}
  \label{eq:5}
||H_{int}\ket{\psi}_{\texttt{SB}}|| \lessapprox \sqrt{\int_0^{\infty} J(\omega) \coth(\beta_{\rm eff} \omega/2)/2},
\end{equation}
where $\beta_{\rm eff}=\frac{1}{k_B T_{\rm eff}}$.
We can evaluate Eq.~(\ref{eq:5}) for the Ohmic case using Mathematica
\begin{equation}
  \label{eq:3}
  ||H_{int}\ket{\psi}_{\texttt{SB}}|| \lessapprox \sqrt{\frac{\alpha}{2}}
  \sqrt{-\omega_c^2+\frac{2\Psi'(\frac{1}{\beta_{\rm eff}\omega_c})}{\beta_{\rm eff}^2}},
\end{equation}
where $\Psi'(x)$ is the first derivative of the digamma function
$\Psi(x)=\frac{\Gamma'(x)}{\Gamma(x)}$ where $\Gamma(x)$ is the
Gamma function. For $\frac{1}{\beta_{\rm eff}\omega_c} \ll 1$ we
do a series expansion and obtain \beq
||H_{int}\ket{\psi}_{\texttt{SB}}|| \lessapprox
\sqrt{\frac{\alpha}{2}}\sqrt{\omega_c^2 +\frac{1}{\beta_{\rm
eff}^2}\left(\frac{\pi^2}{3}+O\left(\frac{1}{\beta_{\rm eff}
\omega_c}\right)\right)}. \label{coolingbound} \eeq Unlike
Eq.~(\ref{hb-bound}), this bound does not involve extensive
quantities, such as the total energy $\langle H_{\texttt{B}}
\rangle_{\psi_{\texttt{SB}}}$ of the bath. However,
Eq.~(\ref{coolingbound}) still involves the high-frequency cut-off
$\omega_c$ because of the zero-point fluctuations of the bath.

\section{Conclusion}
Some important open questions remain in the area of fault-tolerant quantum computation. Most importantly,
is there a threshold result for non-Markovian error models with system-local Hamiltonians and no further
assumptions on the bath? Is this a technical problem, i.e. how can one 
efficiently estimate $Q_B^r$, or are there specific malicious
system-bath Hamiltonians that have such effect that the norm of
the bad faults does not become smaller when increasing $r$? The
next question is whether a better analysis is possible for the
spin-boson model, which is a highly relevant decoherence model. One
would like to evaluate the effect of $H_{\texttt{SB}}$ in the
sector of \emph{physical} states, but the characterization of
these physical states is unclear due to the non-Markovian
dynamics. For real systems one is probably interested in a finite
memory time $\tau > t_0$ which may be simpler to solve. For
example, one can derive the superoperator for a single spin qubit
coupled to a bosonic bath in the Born approximation
\cite{LD-spinbos}, however a derivation involving more than one
system qubit may be too hard to do analytically.

\subsection{Acknowledgements}
We would like to thank David DiVincenzo and Dorit Aharonov for helpful discussions and Andrew Steane for
comments and suggestions. This work was supported
in part by the NSA and the ARDA through ARO contract~No.\ DAAD19-01-C-0056.

\appendix

\section{Bounds on Fault-Path Norms}
\label{appendixf}

\begin{lem}[Fault-Path Norms]
Consider the entire unitary evolution $Q^r$ of a quantum computation on $\texttt{SB}$. Let
$\forall \texttt{q}\in \texttt{S},\;\Delta_{\texttt{SB}}[\texttt{q}] \leq \lambda_0$. We expand $Q^r$ as
a sum over fault-paths which are characterized by a set of faulty locations $\mathcal{I}$.
A fault-path operator with $k$ faults has norm bounded by
\beq
||E(\mathcal{I}_k)|| \leq  ( 2\lambda_0 t_0)^k.
\label{fpboundeq}
\eeq
\label{fpbound}
\end{lem}

\begin{proof}
We do a Trotter-expansion of $Q^r$ and obtain a tree with an
infinite number of branches each of which corresponds to a certain
time-resolved fault-path. Every time a fault occurs at some time
$t_m$ and location $i_m$ we append unitary evolutions for the
remaining time of the location, since we do not care that more
faults occur in that time-interval, the location has failed
anyway. These time-resolved fault-path are characterized by an
index set $\mathcal{T}=((i_1,t_1),(i_2,t_2),\ldots,(i_k,t_k))$ where
$i_1,\ldots,i_k$ is the set of locations of the faults and
$t_1,\ldots,t_k$ label the specific times that the faults occur at
the locations. Every such time-resolved fault-operator with $k$
faults has norm bounded \beq ||E(\mathcal{T}_k)|| \leq \left(\frac{ 2 t
\lambda_0}{n}\right)^k. \eeq
Now we need to group these
time-resolved fault-paths corresponding to faults at sets of locations.
For fixed $n$ faults can occur in time-intervals of length $t/n$
and thus during a time $t_0$ $\frac{t_0 n}{t}$ time-resolved
faults can occur. This implies that \beq ||E(\mathcal{I}_k)|| \leq
||\sum_{\mathcal{T}_k \rightarrow \mathcal{I}_k} ||E(\mathcal{T}_k)|| \leq ( 2 \lambda_0 t_0)^k.
\eeq
\end{proof}


\begin{thebibliography}{AHHH02}

\bibitem{shor-faulttol}
P.~W. Shor.
\newblock Fault-tolerant quantum computation.
\newblock In {\em Proceedings of 37th FOCS}, pages 56--65, 1996.

\bibitem{AB-faulttol}
D.~Aharonov and M.~{Ben-Or}.
\newblock Fault tolerant quantum computation with constant error.
\newblock In {\em Proceedings of 29th STOC}, pages 176--188, 1997,
  \url{http://arxiv.org/abs/quant-ph/9611025}.

\bibitem{klz-faulttol}
E.~Knill, R.~Laflamme, and W.~Zurek.
\newblock Resilient quantum computation: Error models and thresholds.
\newblock {\em Proc. R. Soc. Lond. A}, 454:365--384, 1997,
  \url{http://arxiv.org/abs/quant-ph/9702058}.


\bibitem{AB-ftsiam}
D. Aharonov and M.~Ben-Or.
\newblock Fault-tolerant Quantum Computation with Constant Error Rate.
\newblock To appear in the SIAM Journal of Computation,
\url{http://arxiv.org/abs/quant-ph/9906129}.



\bibitem{gottesman-faulttol}
D.~Gottesman.
\newblock A theory of fault-tolerant quantum computation.
\newblock {\em Phys. Rev. A}, 57:127, 1998,
  \url{http://arxiv.org/abs/quant-ph/9702029}.

\bibitem{preskill-faulttol}
J.~Preskill.
\newblock Fault-tolerant quantum computation.
\newblock In Lo et~al. \cite{book-LPS}, pages 213--269.

\bibitem{KLZ-res}
E.~Knill, R.~Laflamme, and W.~Zurek.
\newblock Resilient quantum computation.
\newblock {\em Science}, 279:342--345, 1998.

\bibitem{steane-overhead}
A.M. Steane.
\newblock Overhead and noise threshold of fault-tolerant quantum error
  correction.
\newblock {\em Phys.\ Rev.\ A}, 68:042322, 2003,
\newblock \url{http://arxiv.org/abs/quant-ph/0207119}.

\bibitem{DB-pureft}
W.~D\"ur and H.-J. Briegel.
\newblock Entanglement purification for quantum computation.
\newblock {\em Phys. Rev. Lett.}, 90:067901, 2003,
  \url{http://arxiv.org/abs/quant-ph/0210069}.

\bibitem{knill-newft}
E.~Knill.
\newblock Scalable quantum computation in the presence of large detected-error
  rates.
\newblock 2003, \url{http://arxiv.org/abs/quant-ph/0312190}.

\bibitem{kitaev-top}
A.~Kitaev.
\newblock Fault-tolerant quantum computation by anyons.
\newblock {\em Annals Phys.} 303:2, 2003;
\newblock \url{http://arxiv.org/abs/quant-ph/9707021}.

\bibitem{alicki+-faulttol}
R.~Alicki, M.~Horodecki, P.~Horodecki, and R.~Horodecki.
\newblock Dynamical description of quantum computing: generic nonlocality of
  quantum noise.
\newblock {\em Phys. Rev. A}, 65:062101, 2002.

\bibitem{fn1}
We could alternatively
use an interaction that involves a `three body' term such as
$\sigma_k[\texttt{q}_i]\otimes \sigma_l[\texttt{q}_j]\otimes A_{kl}$, but this will not make much difference
in the analysis.

\bibitem{fn2}
The freedom is really stronger than this. We
could add a term ${\bf I}_{\texttt{S}}[\texttt{q}] \otimes
O_{\texttt{B}}$ for arbitrary bath operator $O_{\texttt{B}}$
acting on the bath of qubit $\texttt{q}$ since our analysis
(basically Lemma \ref{ebound}) will not depend on the bath
dynamics. It is not clear that this extra freedom makes a big
difference in the analysis. For example, it can be proven that it
does not solve the problems that we face with the norm of the
interaction Hamiltonian for the spin-boson bath, see Section
\ref{strength}.

\bibitem{book-nielsen&chuang}
M.~A. Nielsen and I.~L. Chuang.
\newblock {\em Quantum computation and quantum information}.
\newblock Cambridge University Press, Cambridge, U.K., 2000.

\bibitem{KLV-noise}
E.~Knill, R.~Laflamme, and L.~Viola.
\newblock Theory of quantum error correction for general noise.
\newblock {\em Phys. Rev. Lett.}, 84:2525--2528, 2000,
  \url{http://arxiv.org/abs/quant-ph/9908066}.

\bibitem{book-LPS}
H.-K. Lo, S.~Popescu, and T.P. Spiller, editors.
\newblock {\em Introduction to Quantum Computation}.
\newblock World Scientific, Singapore, 1998.

\bibitem{fn3}
Codes encoding more that one qubit could also be used, but they will complicate the analysis.

\bibitem{alicki:crit}
R. Alicki.
\newblock Comments on "Fault-Tolerant Quantum Computation for Local Non-Markovian Noise".
\newblock 2004, \url{http://arxiv.org/abs/quant-ph/0402139}.


\bibitem{fn4}
We could generalize this to $c$-systemlocal Hamiltonians without much ado.

\bibitem{preskill:faulttol}
J.~Preskill.
\newblock Fault-tolerant quantum computation.
\newblock pages 213--269. World Scientific, Singapore, 1998.

\bibitem{knill:teleft}
E.~Knill.
\newblock Scalable quantum computation in the presence of large detected-error
  rates.
\newblock 2003, \url{http://arxiv.org/abs/quant-ph/0312190}.




\bibitem{wineland+-iontraps}
D.J. Wineland, M.~Barrett, J.~Britton, J.~Chiaverini, B.~DeMarco, W.M. Itano,
  B.~Jelenkovic, C.~Langer, D.~Leibfried, V.~Meyer, T.~Rosenband, and
  T.~Sch\"atz.
\newblock Quantum information processing with trapped ions.
\newblock To appear in Proceedings of the Discussion Meeting on Practical
  Realisations of QIP, held at the Royal Society, 2002,
  \url{http://arxiv.org/abs/quant-ph/0212079}.

\bibitem{STD-local}
K.M. Svore, B.M. Terhal, and D.P. DiVincenzo,
\newblock Local Fault-Tolerant Quantum Computation,
\newblock \url{http://arxiv.org/abs/quant-ph/0410047}.

\bibitem{gottesman-localft}
D.~Gottesman.
\newblock Fault-tolerant quantum computation with local gates.
\newblock {\em Jour. of Modern Optics}, 47:333--345, 2000,
  \url{http://arxiv.org/abs/quant-ph/9903099}.

\bibitem{steane-arch}
A.~Steane.
\newblock Quantum computer architecture for fast entropy exchange.
\newblock {\em Quantum Information and Computation}, 2(4):297--306, 2002,
  \url{http://arxiv.org/abs/quant-ph/0203047}.

\bibitem{fn5}
Of course methods also exist for turning off
unwanted J-couplings by refocusing.


\bibitem{KLG-esd}   
A.~Khaetskii, D.~Loss, and L.~Glazman.
\newblock Electron spin decoherence in quantum dots due to interaction with
  nuclei.
\newblock {\em Phys. Rev. B}, 67:195329, 2003,
  \url{http://arxiv.org/abs/cond-mat/0211678}.

\bibitem{book-weiss}
U.~Weiss.
\newblock {\em Quantum Dissipative Systems}.
\newblock World Scientific, Singapore, 2000.

\bibitem{LD-spinbos}
D.~Loss and D.P. DiVincenzo.
\newblock Exact {B}orn approximation for the spin-boson model.
\newblock \url{http://arxiv.org/abs/cond-mat/0304118}.

\end{thebibliography}

\end{document}